\newif\ifrelease
\releasefalse %
\releasetrue %

\ifrelease
	\documentclass[conference]{IEEEtran}
\else
	\documentclass[conference]{IEEEtran}
\fi

\IEEEoverridecommandlockouts

\usepackage{tikz}
\usepackage{amsmath}
\usepackage{enumitem}
\usepackage{graphicx}
\usepackage{pifont}
\usepackage{threeparttable}

\usepackage{cite}
\usepackage{hyperref}
\usepackage{xspace}
\usepackage{xurl}

\usepackage{filecontents}

\usepackage{array}
\newcolumntype{M}[1]{>{\centering\arraybackslash}m{#1}}
\newcommand{\cmark}{\textcolor{green}{\ding{51}}} %
\newcommand{\xmark}{\textcolor{red}{\ding{55}}}   %

\usepackage[capitalize]{cleveref}
\crefname{figure}{Figure}{Figures} %

\ifrelease
    \usepackage[disable]{todonotes} %
\else
    \usepackage[]{todonotes} %
\fi

\usepackage{soul}

\usepackage{tcolorbox}
\usepackage{xcolor}

\graphicspath{{Figures/}}

\newcommand{\tool}{\textsc{PickleFuzzer}\xspace}

\def\addauthnote#1#2{%
	\expandafter\def\csname#1\endcsname##1{%
		\todo[size=\footnotesize,color=#2,inline]
			{\textbf{\underline{\texttt{#1}}:} ##1}\xspace}
   \expandafter\def\csname#1li\endcsname##1{%
		\todo[size=\footnotesize,color=#2,inline,inlinewidth=5cm, noinlinepar]
			{\textbf{\underline{\texttt{#1}}:} ##1}\xspace}
    \expandafter\def\csname#1Res\endcsname##1{%
		\todo[size=\footnotesize,color=gray,inline]
			{\textbf{\underline{[RESOLVED] \texttt{#1}}:} ##1}\xspace}
}

\addauthnote{andreas}{green!25}
\addauthnote{justin}{blue!25}

\newcommand{\ie}{i.e.,\xspace}

\newcommand{\eg}{e.g.,\xspace}

\newcommand{\etal}{\textit{et al.}\xspace}

\newcommand{\ignore}[1]{}

\newcommand{\summarybox}[1]{%
\begin{tcolorbox}[width=\linewidth, colback=yellow!10!white, top=1pt, bottom=1pt, left=2pt, right=2pt]
#1
\end{tcolorbox}}

\newcommand{\totalErrorsRunA}{13\xspace}

\newcommand{\totalErrorsRunB}{1\xspace}

\newcommand{\totalErrorsFuzzer}{14\xspace}

\newcommand{\totalErrorsManual}{14\xspace}

\newcommand{\totalErrorsOverlap}{10\xspace}

\newcommand{\totalErrorsFuzzerOnly}{4\xspace}

\newcommand{\totalErrorsManualOnly}{4\xspace}

\newcommand{\totalErrorsExceptionType}{13\xspace}

\begin{document}

\date{}

\title{\Large \bf \tool: A Case Study in Fuzzing for Discrepancies \\ Between
Python Pickle Implementations}

\author{
{\rm Justin Applegate}\\
Brigham Young University
\and
{\rm Andreas Kellas}\\
Columbia University
}

\maketitle

\ifrelease
\else
	\thispagestyle{plain}
	\pagestyle{plain}
\fi

\begin{abstract}\label{SEC:ABSTRACT}
Python's native serialization protocol, pickle, is a powerful but insecure format
for transferring untrusted data.
It is frequently used, especially for saving machine learning models, despite
the known security challenges.
While developers sometimes mitigate this risk by restricting imports during
unpickling or using static and dynamic analysis tools, these approaches are
error-prone and depend heavily on accurate interpretations of the Pickle Virtual
Machine (PVM) opcodes.
Discrepancies across Python’s three native PVM modules—\texttt{pickle} (Python),
\texttt{\_pickle} (C), and \texttt{pickletools} (Python disassembler)—can lead
to incorrect detection of malicious payloads and undermine existing defenses.

To efficiently and scalably identify discrepancies, we present \tool, a custom
generation-based fuzzer that identifies inconsistencies across pickle
implementations.
\tool generates pickle objects, passes them to each implementation, and
detects differences in thrown exceptions or changes to key internal states.
\tool generates pickle objects using a grammar, which we developed to account
for the missing pickle specification.
\tool determines discrepancies by comparing the execution behaviors of each
test implementation, rather than requiring a specification-derived oracle.
\tool successfully detected \totalErrorsFuzzer new discrepancies between the
pickle implementations.
Four discrepancies are critical and can be used to bypass security-critical
scanning tools like those deployed on the popular model hosting platform,
Hugging Face.
We disclosed all findings to the Python Software Foundation for remediation,
and additionally disclosed the security issues to a bug bounty platform and
were awarded a \$750 bounty.
We demonstrate that differential testing is a viable approach for identifying
security-relevant discrepancies in important pickle implementations, and our
work can lead to promising future directions for finding deeper pickle bugs with
more directed fuzzing.

\end{abstract}

\section{Introduction}\label{SEC:INTRODUCTION}

Pickle is the native Python serialization interface, and it offers a powerful
and flexible tool for saving and loading data.
This makes it very popular, but also presents many well-studied security
risks~\cite{slaviero_sensepost_2010,slaviero_shellcoding_nodate,pickle-fuzz},
so official documentation warns users to never unpickle untrusted data~\cite{pep307}.
Despite these warnings, pickle remains a popular serialization format for uses
like web applications~\cite{docs_django} and ML model
sharing~\cite{sultanik_dill_2021,kellas_pickleball,splunk}.
To manage these continued risks, security researchers propose new mitigations,
including pickle scanners that analyze pickle objects for the purpose of
identifying malicious pickle files~\cite{modelscan,picklescan,fickling}.
These scanners are the most widely deployed pickle security mitigation, with
integration and investment on platforms like Hugging
Face~\cite{huggingface_protectai,huggingface_protectai_docs,huggingface_pickle},
which host over 2 million machine learning models, many of which use the
pickle format~\cite{splunk,kellas_pickleball}.

To protect users, pickle security tools \textit{must} correctly model
pickle semantics, but this is complicated by the fact that there is no
official pickle specification.
Instead, the pickle serialization interface is implementation defined, with
multiple \textit{de facto} implementations in the CPython release, maintained
officially by the Python Software Foundation~\cite{docs_python_pickle}.
Scanners are left to rely on official implementations like the
\texttt{pickletools} disassembler module, which is one of three pickle parsers
in CPython.
However, the \texttt{pickletools} module semantics can sometimes diverge from
the other two implementations that are used to actually deserialize the object,
creating a gap for attackers to exploit.
An attacker that can craft malicious pickle objects that are parsed differently
during scanning can bypass the security controls and trick users into loading
the malicious object~\cite{reversinglabs:2025:brokenpicklebypass,liu:2025:hideandseek}.

To close these security gaps, we need tools that efficiently identify
discrepancies between pickle implementations.
This is again complicated by the lack of pickle specification, because there is
no single ground truth for comparison.
The solution should be scalable, to consider the multiple existing and future
implementations and versions.
To overcome the challenge of a missing specification, our insight is that pickle
security discrepancies are identifiable by comparing the execution behaviors of
each implementation on the same inputs, as oracles in \textit{differential
testing}.

Using this insight, we designed and implemented \tool, the first automated tool
for testing pickle modules for implementation errors.
\tool is a custom generation-based fuzzer that runs the same pickle object input
through all each of the target implementations of the parsing or unpickling
process and notes any inconsistencies during runtime.
It detects different execution behaviors that are sensitive to (1) errors
occurring and (2) internal storage state at the end of unpickling.

\tool generates test pickle objects and observes the execution of each of the
test pickle implementations.
To generate valid pickle objects, we needed to construct a grammar.
Since there is no specification to consult, we studied the pickle implementations
to categorize the opcodes and produce a generation function.
To identify discrepancies between implementations, we needed to determine properties
of the execution to compare.
We identified error status and internal state as comparison candidates, and
designed \tool to collect these from execution traces for comparisons.

To evaluate \tool and demonstrate its effectiveness at finding differentials
in critical pickle implementations, we used it to fuzz the three pickle
implementations in the CPython release.
\tool identified \totalErrorsFuzzer discrepancies.
Of these, we identified four security critical discrepancies that could permit
an attacker to bypass security scanners with malicious inputs.
We disclosed the security critical discrepancies to the \url{huntr.com} bug
bounty platform and were awarded a \$750 bounty.

\smallskip\noindent\textbf{Contributions:}
\begin{itemize}[itemsep=1pt, topsep=2pt, label={\large\textbf{\textbullet}}]
  \item We created a grammar by manually reviewing existing pickle implementations.
    The grammar serves to generate valid test pickle programs, and can be reused
    for future generative testing in pickle, to help others overcome the lack
    of existing pickle specification.
  \item We designed and implemented \tool, the first automated tool for
  identifying errors in pickle implementations.
  \tool uses our pickle  grammar and applies differential testing to identify
  discrepancies between pickle implementations.%
    \footnote{
      \tool is available at \url{https://github.com/Legoclones/PickleFuzzer}.
    }
  \item We evaluated \tool's ability to identify discrepancies, especially
    security critical ones, in pickle implementations in the CPython release.
    \tool identified \totalErrorsFuzzer, four of which are security critical,
    and were disclosed as issues to the Python Software Foundation.
    To date, six fixes have been released.
\end{itemize}

\section{Background and Related Work}\label{SEC:BACKGROUND}

Here we describe the widespread use of the pickle module and its security risks,
(\S\ref{SEC:BACKGROUND_PICKLE_SECURITY_RISKS}), the dangers of discrepancies
between pickle implementations (\S\ref{SEC:BACKGROUND_PICKLE_DISCREPANCIES}),
and related work that applies differential testing to identify flaws in parsers,
especially those without specifications
(\S\ref{SEC:BACKGROUND_DIFFERENTIAL_TESTING}).

\subsection{Pickle and its Security Risks}\label{SEC:BACKGROUND_PICKLE_SECURITY_RISKS}

Pickle is a highly expressive and convenient serialization format, which
contributes to its widespread adoption, but also to its security risks.
When a Python object is serialized with the pickle module, the object is
converted to a sequence of opcodes that we call (interchangeably) a
\textit{pickle program} or \textit{pickle object}.
When the object is deserialized, the \textit{Pickle Virtual Machine} (PVM)
executes the opcodes to reinstantiate the Python object.
This expressiveness allows programmers to represent nearly any Python object as
a pickle program.
It also gives attackers a powerful primitive for executing arbitrary code; when
they can control the opcodes in a pickle object, they can invoke arbitrary code
during the deserialization
process~\cite{slaviero_sensepost_2010,slaviero_shellcoding_nodate,pickle-fuzz}.
Therefore, users are advised not to unpickle inputs that come from untrusted
sources~\cite{pep307}.

Despite security warnings, pickle is widely used.
Pickle objects were often used to store web cache content served by webservers
developed in frameworks like Django~\cite{docs_django}.
More recently, pickle is used to serialize machine learning models for sharing
on platforms like Hugging Face~\cite{huggingface}.
Sultanik \etal first observed the security risks of the prevalence of pickle in
the ML ecosystem, and released the fickling tool~\cite{sultanik_dill_2021} to
inspect and manipulate pickle-based models.
One study estimated that in 2023, ~80\% of models in the open model ecosystem
are pickle based~\cite{splunk}, and another shows that over a two year period
ending in March 2025, pickle model repository downloads on Hugging Face
\textit{increased} from 500 million downloads per month to 2.1 billion downloads
per month~\cite{kellas_pickleball}.

Because of the known risks of unpickling untrusted data, security researchers
propose secure alternatives.
Early attempts included annotating Python classes with the
\texttt{\_\_safe\_for\_unpickling\_\_} attribute, but this was easily bypassed
and deprecated~\cite{python_commit_16c8}.
Security-conscious developers can write custom unpickling implementations by
overriding specific classes, but Huang \etal showed that 25\% of studied
projects contained insecure custom implementations~\cite{huang_pain_2022}.
In the machine learning setting, restricted loading environments are used to
enforce a subset of domain-specific policies, but they are specialized to
specific libraries~\cite{weights_only_unpickler,kellas_pickleball}.

The most widely deployed ML pickle protection are scanners, which take pickle
objects as inputs and analyze them to determine if they are malicious.
Platforms like Hugging Face integrate scanners directly into their website GUI
to visually alert users when they detect a pickle-based model with malicious
behaviors.
Hugging Face, which hosts 2.7 million models in April 2026, integrates
multiple pickle scanners into its platform to alert users when a model appears
malicious%
~\cite{huggingface_pickle,huggingface_protectai,huggingface_protectai_docs}.

\subsection{Dangers of Pickle Implementation Discrepancies}\label{SEC:BACKGROUND_PICKLE_DISCREPANCIES}

Pickle is an implementation-specified language with no official specification,
but rather three \textit{de facto} implementations maintained in the core
CPython project.
Some implementations are pickle \textit{deserializers}, which parse and execute
pickle object opcodes in the PVM, while other implementations are pickle
\textit{disassemblers}, which parse the opcodes for analysis.
The original pickle deserializer implementation is the \texttt{pickle} module
written in Python~\cite{python_github_picklepy}.
Another deserializer implementation is the \texttt{\_pickle} module (previously
known as \texttt{cPickle}), which is a C version implemented for
performance-optimized deserialization~\cite{python_github_picklec}.
Typically, the \texttt{pickle.loads()} method first attempts to load from the
optimized \texttt{\_pickle} module before using the regular \texttt{pickle}
implementation~\cite{python_github_picklepy}.
Unlike the deserializers, the \texttt{pickletools} module is a disassembler and
library for external tools to use when parsing pickle
objects~\cite{python_github_pickletools}.
Security critical applications that parse pickle objects, like
Picklescan~\cite{picklescan} and ModelScan~\cite{modelscan}, depend on \texttt{pickletools}
for correct pickle parsing in order to identify malicious pickle objects.
Future applications may yet implement their own pickle parsing or deserializing
logic, but do not have a specification to refer to for correctness.

Discrepancies between pickle parser implementations, whether in deserializer or
disassembler modules, can have disastrous security effects.
Consider two pickle parsers: one is used as a disassembler for a security tool
to scan pickle objects before deserialization, and the other is used to actually
deserialize the objects once deemed safe.
When the two parsers differ for the same input pickle object, the security
tool may predict that the object is benign, but the deserialization module
executes the object's malicious behavior.
Recent work by Liu \etal~\cite{liu:2025:hideandseek} describes
``Exception-Oriented Programming'' as a technique to bypass pickle scanners,
where an attacker crafts a malicious pickle object that parses and executes
correctly in the \texttt{pickle} module, but which results in an exception when
scanned by the security scanner.
Scanners, like those deployed on Hugging Face, will fail to alert on malicious
pickle objects, putting users at risk.
Researchers have already discovered malicious pickle-based models on Hugging
Face that abuse this exact
scenario~\cite{reversinglabs:2025:brokenpicklebypass},

Despite pickle opcodes appearing simple, the differences between pickle parsers
can manifest in different ways, including in how opcodes interact with their
parameters and the internal structures of the pickle machine.
Pickle instructions appear simple because pickle programs execute in straight
lines, with no branching instructions, returning the object at the top of the
execution stack when the program halts.
However, the pickle machine internally maintains three main storage areas for
opcodes to interact with: the \textit{stack}, \textit{metastack}, and
\textit{memo}.
The stack is the main storage area, and approximately half of the opcodes are
used to push new data to the stack.
The metastack is used to segment the stack into logical frames, where opcodes
can push and pop data all at once between designated markers.
The memo is a dictionary where values can be stored (associated with a numerical
key) and referenced later, ensuring that these values only need to be
deserialized once.
Further, the pickle deserialization process also takes some optional arguments,
including an \textit{encoding} and an \textit{out-of-band buffer}.
When opcodes decode input bytes into Python strings, the \textit{encoding}
parameter specifies the encoding format used, such as UTF-8 or Latin-1.
Out-of-band buffers allow the pickle provider and pickle consumer to share a
large list of data beforehand, and during the unpickling process the pickle can
simply ask for the next item in the pre-shared out-of-band buffer and then use
it accordingly \cite{docs_python_pickle}.
Each pickle version must carefully implement its opcodes to interact
consistently with these features that affect the internal execution state during
deserialization.

\subsection{Differential Testing}\label{SEC:BACKGROUND_DIFFERENTIAL_TESTING}

\cref{SEC:BACKGROUND_PICKLE_DISCREPANCIES} discusses how discrepancies
between pickle implementations can have real security risks, motivating the need
to find such discrepancies at scale.
Previous work applies \textit{differential testing} to identify discrepancies
between implementations of other parser types.
Differential testing applies a single test input to two or more implementations
of a protocol or parser~\cite{McKeeman1998DifferentialTF}.
If any implementation produces a different result when compared to others, a
potential bug, or differential, is exposed.

Recent work shows that differential testing is promising for exposing security
flaws in systems that are difficult to specify, or that lack specifications.
Assemblers often accept different grammars and have no specified grammar. Kim
\etal~\cite{asfuzzer} developed a fuzzer that utilizes error messages to infer underlying
grammar rules and compares output across 4 real-world assemblers.
They successfully discovered 497 parser discrepancies across 6 popular
architectures.
URLs are defined by two different standards---namely IETF RFC 3986 and the
WHATWG URL Living Standard---but URL parsing implementations often try to
conform to both standards, or comply with obsolete RFCs.
Kallus and Smith designed a differential fuzzer for Python URL parsing
implementations using grammar-based mutations that discovered several previously
unknown bugs from multiple implementations~\cite{kallus_dippygram}.
Like pickle scanners, email MIME parsers are used to identify malicious
intent, and Andarzian \etal~\cite{andarzian:2025:mimedifferentialfuzzing}
demonstrate the potential for differential testing to identify discrepancies
in this setting.
Reynolds \etal~\cite{equivocal_urls} tested fifteen different URL parsers for discrepancies and grouped them into seven main pitfalls, demonstrating how the discrepancies can be used to bypass URL scanners.
Our work is the first to apply differential testing to automatically test
pickle implementations, and must overcome pickle-specific design challenges.

\section{\tool Design}\label{SEC:OVERVIEW}

The previous section describes how implementation differences between pickle
modules can manifest in security risks, and identifies differential testing
as one technique that has been used to solve similar problems.
We recognize the need for an automated tool to identify pickle module
discrepancies, which we realize with our system \tool.
Here, we discuss the design of \tool by defining the problem that it aims to
solve (\S\ref{SEC:OVERVIEW_PROBLEM_DEFINITION}), describing the system design
(\S\ref{SEC:OVERVIEW_DESIGN}), and discussing its design limitations
(\S\ref{SEC:OVERVIEW_LIMITATIONS}).

\subsection{Threat Model and Solution Requirements}\label{SEC:OVERVIEW_PROBLEM_DEFINITION}

\smallskip\noindent\textbf{Threat Model.}
In this work, we are chiefly concerned with the attacker who abuses the
differences between pickle implementations to create a malicious pickle object
that results in different analysis or deserialization outcomes.
The attacker then uses the different outcomes to achieve their malicious goal,
\eg by causing a third-party security scanner to misclassify the malicious
pickle object, leading the application to unpickle the malicious object with a
different outcome.
As we discuss in \cref{SEC:BACKGROUND}, this is an important adversary model
due to pickle's prevalence and the reliance on pickle security scanners in the
open model ecosystem.
We also aim to identify less security-critical discrepancies between pickle
implementations in order to improve unpickling reliability.

\smallskip\noindent\textbf{Solution Requirements.}
In order to prevent attackers from abusing discrepancies in the different pickle
implementations, we require a system that can reliably and scalably find
differences between them.
The system must precisely identify discrepancies with minimal false positives
(incorrectly identified discrepancies) and false negatives (missed
discrepancies).
The system must be scalable to quickly analyze the current and future versions
of pickle implementations in order to identify discrepancies before attackers.

\subsection{\tool Design}\label{SEC:OVERVIEW_DESIGN}

We designed \tool in an attempt to meet the requirements described in
\cref{SEC:OVERVIEW_PROBLEM_DEFINITION}.
\tool is a differential fuzz testing system that identifies discrepancies
between a set of pickle implementations.

\begin{figure*}
  \begin{center}
  \includegraphics[width=0.9\textwidth]{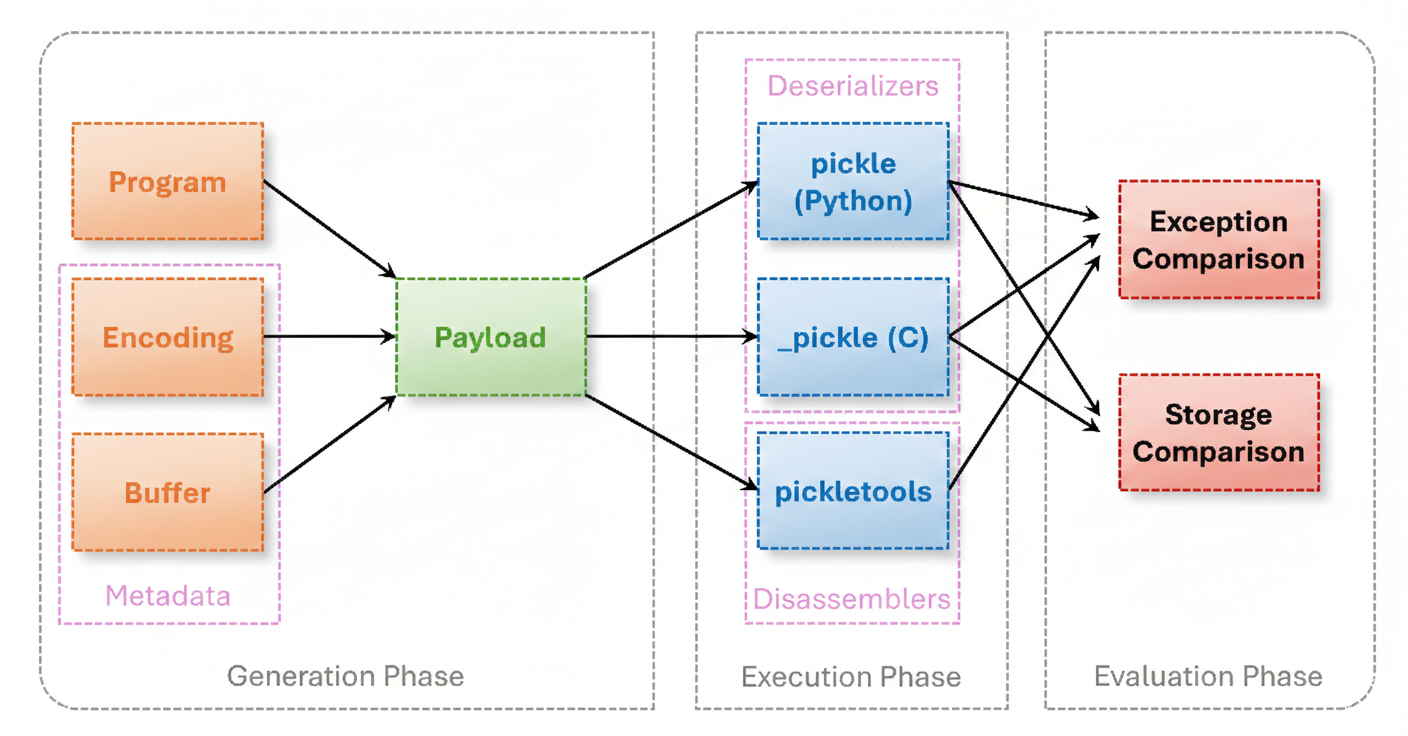}
  \caption{\label{fig:workflow} Each iteration of \tool has three phases.
  In the Generation Phase, a pickle program is generated from a grammar and
  combined with randomly selected encoding and buffer metadata to create a test
  payload.
  In the Execution Phase, the test payload is run by each of the test modules,
  and execution trace information is collected.
  In the Evaluation Phase, the execution information is compared to identify
  discrepancies.
  When a module is a disassembler, like \texttt{pickletools}, it's execution
  information is only used in exception comparisons, while deserializer modules
  will pass execution information to be checked for both kinds of comparisons.
  }
  \end{center}
\end{figure*}

At a high level, \tool iteratively generates pickle objects and evaluates
whether each pickle implementation parses the objects in the same manner.
Each iteration is a sequence of three phases: \textit{generation},
\textit{execution}, and \textit{evaluation} (see~\cref{fig:workflow}).
In the Generation Phase, \tool generates a test pickle object and combines it
with additional test metadata to create a test \textit{payload}.
In the Execution Phase, \tool passes the payload to the multiple target pickle
implementations under test, in a standardized environment, and produces an
execution result.
In the Evaluation Phase, \tool compares the execution result of each
implementation to identify any discrepancies between them, which are saved.
At the end of each iteration, \tool begins again at the Generation Phase to
create a new test input payload.
This proceeds until \tool reaches its user-designated fuzzing time limit.

\tool identifies discrepancies based on execution behaviors that, for a fixed
valid pickle object, differ in (1) the exceptions raised, or (2) the internal
storage contents of the pickle machine.
\tool is not designed to identify discrepancies that result from malformed
pickle objects, arbitrarily sized objects, or that result in different execution
state differences.
As a fuzzer, \tool is also not expected to prove that there are no
discrepancies, but rather to efficiently identify the ones that do exist.

\subsubsection{Generation Phase}\label{SEC:OVERVIEW_DESIGN_GENERATION}

The purpose of the Generation Phase is to generate a test input, or
\textit{payload}.
The payload must contain (1) a valid pickle object with opcode instructions to
test, and (2) any configuration metadata that can influence the parsing or
deserializing behavior of a pickle module.
This metadata includes \textit{encoding} and \textit{out-of-band buffer}
configuration settings, which determine parsing behaviors of the module.

\smallskip\noindent\textbf{Opcode Generation.}

The payload must contain a valid pickle program composed of opcode
instructions, which \tool creates with a grammar-based generation function.%
    \footnote{
      To see more details about the pickle format and opcodes, see~\cite{pickledoc}.
    }
The generation function generates a random number of opcodes, bounded by a
user-configured upper limit (defaults to 20), and appends the \texttt{STOP}
opcode that demarcates the end of a valid pickle program.

Because there is no specification for pickle opcodes, we studied the three
\textit{de facto} implementations and categorized opcodes for relevant handling
in the generation function:
\begin{enumerate}
  \item opcodes with no arguments;
  \item opcodes with fixed-length arguments;
  \item opcodes with one fixed-length argument that specifies the length of a
    second variable-length argument; and
  \item opcodes with variable-length arguments that are character delimited,
    usually with a newline character.
\end{enumerate}
The generation function specifies each opcode category to correctly generate
valid instructions with randomly instantiated, but valid, arguments.

\smallskip\noindent\textbf{Configurable Metadata.}
Pickle modules permit users to configure certain behaviors of the module when
it is instantiated (prior to deserializing any data), and these configuration
behaviors should also be consistent between pickle module implementations.
We analyzed the \texttt{Unpickler} instantiation options of the class C and
Python deserialization modules to determine configuration options that should
be tested for consistent behaviors between the modules.
We identified the \texttt{encoding} and \texttt{buffers} optional arguments as
potential configurable parameters that need to be handled consistently.
\tool varies these configurable parameters by randomly generating an option for
each during payload generation.

The \texttt{encoding} argument specifies how a deserialization module should
decode non-printable bytes into strings during the unpickling process.
For example, an \texttt{Unpickler} instantiation that is configured to use UTF-8
encoding will treat all string arguments to pickle opcodes as UTF-8 encoded strings.
\tool randomly selects one of four valid encoding arguments for each generated
payload: UTF-8, UTF-16, ASCII, and Latin-1.

The \texttt{buffers} argument lets the \texttt{Unpickler} define sources of
out-of-band buffers that are used during deserialization, often for the purpose
of enabling zero-copy unpickling of large objects~\cite{docs_python_pickle}.
Without the \texttt{buffers} argument set, pickle objects are treated as
self-contained, with all data needed for deserialization contained within the
object.
When the \texttt{buffers} argument is provided, it should be an iterable of
Python objects that are returned sequentially whenever a specific pickle opcode
request an object from the buffers.
\tool randomly selects an initial value for \texttt{buffers} from a set of
possible values, including \texttt{None}, an empty list, a list containing
various Python objects, and other various Python data types.
This allows \tool to explore the semantics of opcodes that interact with the
out-of-band buffers.

\subsubsection{Execution Phase}

The purpose of the Execution Phase is to pass the generated payload to each
of the test modules for parsing or deserialization, and to record the execution
state for later analysis in the Evaluation Phase.
Put simply, the Execution Phase executes the payload using the loading or
parsing API of each pickle module and records the execution state of
the Pickle Machine implementation once the payload is processed.
\tool is designed to overcome two challenges in this phase: (1) exposing the
internal execution state after parsing the payload, and (2) maintaining an
identical execution environment for each module.

The results of the execution state are provided to the Evaluation Phase, and
take one of two forms:
\begin{enumerate}
  \item \textbf{Exception-based state} records whether an exception was raised during
    execution.
  \item \textbf{Storage-based state} records the internal storage resulting from
    execution.
\end{enumerate}
\tool tracks the exception-based state of all test modules, but tracks the
storage-based state for only pickle deserialization modules.
\tool does not track internal storage for disassemblers, because tools like \texttt{pickletools} only parse opcodes rather than executing the pickle program.

\smallskip\noindent\textbf{Execution State Tracking.}
Normally, when a pickle module finishes deserialization or disassembling a
pickle object, it returns limited information about the execution state, but
\tool enriches the module to return additional state information.
In particular, \tool's analysis is enriched by exposing the full contents
of the three storage areas: the stack, metastack, and memo (discussed in
\cref{SEC:BACKGROUND_PICKLE_DISCREPANCIES}).
To make this information accessible, \tool requires the user to apply minor
patches to the target deserialization modules so that the storage areas
are returned as a tuple when deserialization halts.
These modifications do not need to be made in disassembler pickle modules
because the module storage information is not tracked in disassemblers.

\smallskip\noindent\textbf{Identical Execution Environments.}
The Python runtime environment must remain the same across all pickle
deserializations to ensure accurate results.
This applies to different implementations deserializing the same pickle object
as well as the same implementation deserializing different pickle objects.

Some pickle opcodes can import and invoke Python objects in the pickle virtual
environment, which introduces external state effects to the fuzzing execution.
A naive fuzzing implementation could work around this by removing these
importing opcodes from the generation grammar, but this would leave an important
set of pickle behaviors untested.
Instead, \tool strikes a compromising balance by hooking the import opcodes to
perform one of three actions: (1) import a function that does nothing, (2)
import a class with no attributes, or (3) import an instance of the class
defined in (2).
This hook is inserted by patching the pickle module's \texttt{find\_class}
method, which is invoked when an opcode imports a Python module.
One of the three actions is selected by applying a fast hashing function to the
arguments of \texttt{find\_class} to deterministically but randomly return one
of the three actions.
This ensures that opcodes that import modules are still tested, but that the
environment remains consistent for all executions.

To further enable consistent execution environments, \tool is designed to
execute within a Docker container environment that fixes the OS and Python
versions for repeatable results.
We describe the precise configuration of our container environment in
\cref{SEC:IMPLEMENTATION} when we discuss \tool's implementation.

\subsubsection{Evaluation Phase}\label{SEC:OVERVIEW_DESIGN_EVALUATION}

The purpose of the Evaluation Phase is to identify discrepancies between the
execution records of any test modules when executed with the same test input.

\tool identifies two kinds of discrepancies between pickle implementations:
\textit{error discrepancies} and \textit{storage discrepancies}.
Error discrepancies occur when one or more implementations raise an exception
when unpickling a bytestream, but at least one other implementation does not.
Error discrepancies are checked for between all test modules, including
disassemblers and deserializers.
Storage discrepancies occur when the contents of or order of data in any of the
three storage areas differ between pickle implementations.
Storage discrepancies are only checked for between deserialization modules,
since disassembler modules like \texttt{pickletools} do not maintain internal
storage because they do not execute the pickle program.

Other kinds of discrepancies, like time- and resource-based discrepancies, are
not considered because of inherent differences between implementations, such as the C
\texttt{\_pickle} module that was created specifically to be faster and use
fewer resources~\cite{docs_python_pickle}.
We discuss limitations with respect to discrepancy types in~\cref{SEC:OVERVIEW_LIMITATIONS}.

Once discovered, error discrepancies are triaged for deduplication with a simple
analysis that checks whether the given error discrepancy
has already been identified, and if so, discards the saved discrepancy.
This ultimately results in a list of unique exception-based discrepancies identified by \tool. Manual triaging is required to deduplicate storage discrepancies and identify unique differences.

\smallskip\noindent\textbf{Error Discrepancies.}
During the execution result on a test payload, one or more modules may raise
an exception while another does not.
This indicates that there is disagreement between the implementations about what
a valid pickle program is.
All test modules are checked for error discrepancies.

\tool's error identification is coarse-grained and does not consider the
exception types.
If \tool did consider exception types, it may be overwhelmed with false-positive
results;
  so far, little effort has been made in the CPython codebase to standardize the
  exceptions thrown between the implementations, even for the same runtime error%
  ~\cite{python_issue_cpickle}.
Therefore, we designed \tool to consider only that an exception was raised,
rather than its type.

\smallskip\noindent\textbf{Storage Discrepancies.}
Deserialization modules are checked for storage discrepancies.
Storage discrepancies represent differences in the internal pickle machine
memory state during object deserialization, and can appear as differences in
any of the stack, metastack, or memo storage areas.
Each of these storage areas are provided by the Execution Phase for each
execution, and represent all variable names and values that were assigned during
deserialization.
However, storage discrepancies are more challenging to identify than error
discrepancies because of nuance in how Python implements the equality
comparison.
Due to how Python handles the \texttt{==} operator (value comparison) and
the \texttt{is} keyword (object memory comparison), as well as type juggling
in ``truthy'' and ``falsy'' comparisons, we first cast all values to strings
and then compare with the \texttt{==} operator.
For the cast approach to work, we must also discard any addresses included in
variable or string representations, because address space layout randomization
(ASLR) can result in different address names for the same logical value.

\subsection{\tool Limitations}\label{SEC:OVERVIEW_LIMITATIONS}

\tool is designed to solve the requirements identified in
\cref{SEC:OVERVIEW_PROBLEM_DEFINITION}, but it makes some compromises
that introduce limitations, which we describe here.

\smallskip\noindent\textbf{Fuzzing and False Negatives.}
As a fuzzer, \tool is designed to identify positive examples of discrepancies
between pickle implementations, but is not guaranteed (nor expected) to be able to
identify \textit{all} instances of discrepancies in pickle implementations.
This is an acceptable tradeoff in the design of \tool given the scalability
and effectiveness of fuzzing and differential testing for discovering bugs.
The evaluation of \tool (\S\ref{SEC:EVALUATION}) shows empirically that it can
discover impactful instances of bugs, validating our approach.

\tool is a grammar-based generation fuzzer, which limits the kinds
of inputs that it generates.
While grammar-based fuzzing is effective for generating highly structured
program inputs, it also has its own drawbacks compared to mutation-based or
coverage-guided fuzzing~\cite{mallissery_demystify_2023}.
\tool's payloads will always contain well-formed pickle programs that conform
to the grammar, with valid opcodes and arguments.
Specifically, some bugs associated with parsing the syntax may be missed, and if
the grammar is incomplete or incorrect the fuzzer may miss out on available
functionality~\cite{salls_token-level_2021}. See~\cref{SEC:DISCUSSION_ALTERNATE} for further discussion on fuzzing design implications.

\smallskip\noindent\textbf{Discrepancy Types.}
\tool identifies specific kinds of discrepancies based on the result of
executing each module, but it may miss discrepancies of other kinds.
The current design of \tool checks for the presence of exceptions and for
differences in storage contents, and can feasibly be extended for other
relevant kinds of discrepancies.
As we explain in \cref{SEC:OVERVIEW_DESIGN_EVALUATION},
we intentionally exclude time- and resource-based discrepancies and execution
type.
Our evaluation (\S\ref{SEC:EVALUATION}) validates that our decisions to track
exceptions and internal storage state are appropriate for identifying meaningful
discrepancies.
If other relevant discrepancy types are discovered in the future, \tool can
be extended in its Execution and Evaluation phases to incorporate them.

\smallskip\noindent\textbf{Data Length Limits.}
  \tool must balance exploration with reasonable system resource utilization,
  leading to limits on finding discrepancies reached by arbitrarily large
  payloads.
Pickle objects can contain an arbitrarily large number of opcode instructions,
and some opcodes can have arbitrarily large variable-length arguments.
However, permitting \tool to generate arbitrarily-sized pickle programs or
arguments results in high memory and hard-drive resource utilization.
To mitigate this, \tool has a user-defined upper limit on generated payload
sizes, as discussed in \cref{SEC:OVERVIEW_DESIGN_GENERATION}.
\tool will not be able to identify discrepancies that are reached by inputs
with sizes that exceed these limits.
However, these limits are configurable, so a user with more available resources
can still explore larger pickle inputs.

\section{\tool Implementation}\label{SEC:IMPLEMENTATION}

\tool is implemented as a Python application in approximately 1,400 lines of
code (LoC).
It runs in an Ubuntu 24.04 Docker container with Python version 3.13.0.
Each test module is patched to integrate it into \tool, and the patches are
approximately 150 LoC total, in Python and C.
The patches are meant to return additional information about the execution state
after deserialization, remove unnecessary I/O, and overload the
\texttt{find\_class()} as described in \cref{SEC:OVERVIEW_DESIGN}.
The patches are minimal and can be easily extended to new Python versions.

\section{Evaluation}\label{SEC:EVALUATION}

We evaluated \tool with three Research Questions:

\begin{itemize}

  \item \textbf{RQ1: Finding Discrepancies.} Does \tool identify
    discrepancies between pickle implementations?

  \item \textbf{RQ2: Security Impact.} Can a malicious actor use identified
    discrepancies to bypass pickle scanner defenses?

  \item \textbf{RQ3: Search Efficiency.} Is \tool more efficient at identifying
    discrepancies than manual analysis?

\end{itemize}

\subsection{Evaluation Targets and Setup}\label{SEC:EVALUATION_SETUP}

\subsubsection{Target Selection}
\tool is meant to identify discrepancies between implementations of pickle
parsers and deserialization modules.
We evaluated \tool by selecting the three \textit{de facto} pickle
implementations that are distributed in the core CPython modules: the Python
\texttt{pickle} module, the C \texttt{\_pickle} module, and the
\texttt{pickletools} module.
They are each used regularly in critical applications, like web application
frameworks~\cite{docs_django} and ML applications, as well as in security
scanners~\cite{huggingface_protectai,huggingface_pickle,fickling}.
The module versions were all from the CPython 3.13.0 release, and patched in
order to be integrated into the \tool workflow (described in
\cref{SEC:OVERVIEW_DESIGN} and \cref{SEC:IMPLEMENTATION}).

\subsubsection{Environment Setup}
All experiments are conducted on Debian Linux with 32 GB of RAM and 12 i5-10500H
CPU cores @ 2.50GHz.
We permit \tool to use a maximum of 12GB of RAM and six workers.
\tool ran in a Docker container environment with Ubuntu 24.04.

\subsection{RQ1: Finding Discrepancies}\label{SEC:EVALUATION_FINDING_DISCREPANCIES}

\begin{table*}[!tp]
  \centering
  \caption{\tool Discrepancy Results. This table lists every discrepancy discovered
  by either \tool or manual analysis (\S\ref{SEC:EVALUATION_SEARCH_EFFICIENCY}).
  ``Differing Module'' indicates which implementation different from the others under test.
  ``Security Impact'' indicates whether the discrepancy can be used to bypass scanners (\S\ref{SEC:EVALUATION_SECURITY_IMPACT}).
  ``Fixed'' indicates the status of the issue used to track the discrepancy after disclosure.
  }
  \vspace{0.5em}
  \label{tab:discrepancy_summary_long}
  \renewcommand{\arraystretch}{1.3}
  \begin{tabular}{|M{1.4cm}|M{4.8cm}|M{1.2cm}|M{1.2cm}|M{1.4cm}|M{1.9cm}|M{0.9cm}|M{0.9cm}|M{0.65cm}|}
    \hline
    \textbf{Discrepancy Number} & \textbf{Discrepancy Description} & \textbf{Found manually} & \textbf{Found by Fuzzing} & \textbf{Discrepancy Type} & \textbf{Differing Module} & \textbf{Security Impact} & \textbf{Python Issue \#} & \textbf{Fixed}\\
    \hline
    1  &
        \texttt{INT} and \texttt{LONG} opcodes use base 0 in all implementations except \texttt{pickletools}, which uses an explicit base of 10 &
        \cmark & \cmark &
        Error & \texttt{pickletools} &
        \cmark &
        \href{https://github.com/python/cpython/issues/126992}{126992} &
        Fixed  \\
    \hline
    2  &
        All newline-terminated opcode arguments are ended early by null bytes in \texttt{\_pickle}, but throw errors in the other implementations &
        \cmark & \cmark &
        Error & \texttt{\_pickle} &
        \cmark &
        \href{https://github.com/python/cpython/issues/126996}{126996} &
        Open  \\
    \hline
    3  &
        Lack of \texttt{pickletools} encoding support leads to invalid transformation of bytes to strings for some opcodes &
        \cmark & \cmark &
        Error & \texttt{pickletools} &
        \cmark &
        \href{https://github.com/python/cpython/issues/126997}{126997} &
        Fixed  \\
    \hline
    4  &
        \texttt{pickletools} errors if there's any number of items left on the stack when reaching the \texttt{STOP} opcode, while other implementations do not &
        \cmark & \cmark &
        Error & \texttt{pickletools} &
        \xmark &
        \href{https://github.com/python/cpython/issues/127079}{127079} &
        Not Fixed  \\
    \hline
    5  &
        \texttt{pickletools} errors if any memo key is set multiple times, while other implementations don't &
        \cmark & \cmark &
        Error & \texttt{pickletools} &
        \cmark &
        \href{https://github.com/python/cpython/issues/123309}{123309}\textsuperscript{1} &
        Fixed  \\
    \hline
    6  &
        In some cases, the \texttt{INT} opcode of \texttt{\_pickle} may decode an argument as a boolean \texttt{True}/\texttt{False} instead of 1/0 &
        \cmark & \cmark &
        Storage & \texttt{\_pickle} &
        \xmark &
        \href{https://github.com/python/cpython/issues/135241}{135241} &
        Fixed  \\
    \hline
    7  &
        If the reported length of a \texttt{BINSTRING} opcode's argument is greater than 0x80000000, then \texttt{\_pickle} treats the length as positive and valid, while the others treat it as negative and invalid  &
        \cmark & \xmark &
        Error & \texttt{\_pickle} &
        \cmark &
        \href{https://github.com/python/cpython/issues/135321}{135321} &
        Fixed  \\
    \hline
    8  &
        \texttt{pickle} errors if an \texttt{APPENDS} or \texttt{ADDITEMS} opcode immediately proceeds a \texttt{MARK} object on the stack, while the others do not &
        \cmark & \cmark &
        Error & \texttt{pickle} &
        \xmark &
        \href{https://github.com/python/cpython/issues/135573}{135573} &
        Fixed  \\
    \hline
    9  &
        \texttt{PUT} opcode argument is restrained to fit inside an \texttt{ssize\_t} variable in \texttt{\_pickle} but not in other implementations &
        \cmark & \xmark &
        Error & \texttt{\_pickle} &
        \xmark &
        \href{https://github.com/python/cpython/issues/144410}{144410} &
        Not Fixed  \\
    \hline
    10 &
        Large memo indices in two opcodes of \texttt{\_pickle} cause Python to throw an Out-Of-Memory error &
        \cmark & \cmark &
        Error & \texttt{\_pickle} &
        \xmark &
        \href{https://github.com/python/cpython/issues/115952}{115952} &
        Fixed  \\
    \hline
    11 &
        \texttt{\_pickle} checks that a \texttt{NEWOBJ} or \texttt{NEWOBJ\_EX} argument is a tuple, while other implementations accept any iterable &
        \cmark & \cmark &
        Error & \texttt{\_pickle} &
        \xmark &
        \href{https://github.com/python/cpython/issues/135579}{135579} &
        Not Fixed  \\
    \hline
    12 &
        Any whitespace in the \texttt{FLOAT} opcode argument causes \texttt{\_pickle} to error but not other implementations &
        \xmark & \cmark &
        Error & \texttt{\_pickle} &
        \xmark &
        \href{https://github.com/python/cpython/issues/135580}{135580} &
        Open  \\
    \hline
    13 &
        In \texttt{\_pickle}, the \texttt{BUILD} opcode's \texttt{state} argument checks for \texttt{Py\_None}, while others check if it's falsy &
        \xmark & \cmark &
        Error & \texttt{\_pickle} &
        \xmark &
        \href{https://github.com/python/cpython/issues/128965}{128965} &
        Open  \\
    \hline
    14 &
        In the \texttt{FRAME} opcode, \texttt{\_pickle} checks to see that \texttt{n} bytes exist in the stream, but others check to see that \textit{up to} \texttt{n} bytes exist &
        \cmark & \cmark &
        Error & \texttt{\_pickle} &
        \xmark &
        \href{https://github.com/python/cpython/issues/128853}{128853} &
        Not Fixed  \\
    \hline
    15 &
        Splitting an opcode and its argument across multiple frames errors out in \texttt{pickle} but not others &
        \cmark & \xmark &
        Error & \texttt{pickle} &
        \xmark &
        \href{https://github.com/python/cpython/issues/128853}{128853} &
        Not Fixed  \\
    \hline
    16 &
        Frames are not allowed to overlap in \texttt{pickle} but are in \texttt{\_pickle} &
        \cmark & \xmark &
        Error & \texttt{pickle} &
        \xmark &
        \href{https://github.com/python/cpython/issues/128853}{128853} &
        Not Fixed  \\
    \hline
    17 &
        \texttt{\_pickle} checks that the \texttt{REDUCE} argument \texttt{args} is a tuple, while other implementations accept any iterable &
        \xmark & \cmark &
        Error & \texttt{\_pickle} &
        \xmark &
        \href{https://github.com/python/cpython/issues/144412}{144412} &
        Open  \\
    \hline
    18 &
        \texttt{\_pickle} checks that the \texttt{BUILD} argument \texttt{slotstate} is a dictionary, while other implementations only check to see if its falsy &
        \xmark & \cmark &
        Error & \texttt{\_pickle} &
        \xmark &
        \href{https://github.com/python/cpython/issues/144411}{144411} &
        Open  \\
    \hline
  \end{tabular}
  \begin{tablenotes}
    \footnotesize {
        \item{1: This issue was independently discovered and reported by a
        third-party.}
    }
  \end{tablenotes}
\end{table*}

\tool is designed to identify discrepancies between pickle implementations.
To evaluate its ability to do so, we conduct a fuzzing campaign using \tool
on the three pickle modules discussed in \cref{SEC:EVALUATION_SETUP}.

\subsubsection{Methods}

In order for \tool to identify discrepancies between the test modules, we
conducted a fuzzing campaign of 168 hours (one week) using the default resource
limits and settings.
This includes limiting the number of opcodes per payload to between 1 and 20,
setting the number of random items in the buffers argument to 3, setting the
maximum number of digits in unlimited-length ASCII number opcodes to 10 (e.g.,
\texttt{1234567890}), setting the maximum number of bytes in unlimited-length
byte opcodes to 300, and artificially limiting the \texttt{PUT} and
\texttt{LONG\_BINPUT} opcodes to indices of up to 999999 and 65535, respectively, to prevent Out-Of-Memory errors.

In order to try to also identify discrepancies that require large inputs, we
conducted a second fuzzing campaign with relaxed limits on input sizes but for a
more limited 12 hours, to avoid resource exhaustion.
In this run, the maximum number of digits in unlimited-length ASCII number
opcodes was changed to 25 (which can represent numbers larger than the greatest
unsigned 64-bit value), the maximum number of bytes in unlimited-length byte
opcodes was changed to 0x180000000 (6 GB in size, ensuring the maximum was a
number that had a 1/3 chance of being greater than 32 bits), and the
\texttt{PUT} and \texttt{LONG\_BINPUT} opcodes were unrestricted.

\subsubsection{Results}

During the first (week-long) campaign, \tool discovered over 42 million error
discrepancy payloads and 1500 storage discrepancy payloads, which were
triaged down to \totalErrorsRunA total discrepancies.
All \totalErrorsRunA discrepancies were found within the first 6
minutes of the run, half of which were found within 2 seconds.
This demonstrates that \tool is able to quickly identify new discrepancies
between multiple pickle implementations.

In the second (12 hour-long) campaign, \tool discovered \totalErrorsRunB additional
discrepancy, within 10 minutes of the run.
This demonstrates that even though \tool is limited in its ability to generate
arbitrarily large inputs to find discrepancies and handle resource-heavy inputs, increasing its constraints
can still result in new discrepancies found.

Across both runs, \tool discovered \totalErrorsFuzzer discrepancies.
All discrepancy-triggering payloads contain between one and five
opcodes, with an average of 2.4 opcodes.
Only one discrepancy (\#6) was a storage discrepancy, while the other
\totalErrorsExceptionType were error discrepancies.
Across all discrepancies, 1 (7\%) of the differences resulted from the Python
\texttt{pickle} module differing from the other test modules, 9 (64\%) from the
C \texttt{\_pickle} differing, and 4 (29\%) from \texttt{pickletools} differing.
\cref{tab:discrepancy_summary_long} provides a detailed breakdown.

\summarybox{\textbf{RQ1 Summary:} \tool successfully identifies discrepancies between pickle modules. It identifies \totalErrorsFuzzer discrepancies, all within 10 minutes.}

\subsection{RQ2: Security Impact}\label{SEC:EVALUATION_SECURITY_IMPACT}

\tool records all discrepancies that it discovers, without considering
their severity as security vulnerabilities.

However, some discrepancies have a high security impact when they permit attackers
to bypass security scanners and execute malicious code.
We measure \tool's ability to identify impactful discrepancies by reviewing
the \totalErrorsFuzzer discrepancies that it discovered.
We assess whether a discrepancy is a security vulnerability by considering
whether it can be used to bypass security tools that try to identify malicious
pickle objects, and use bug bounty responses for additional validation.

\subsubsection{Methods}
All public pickle-scanning tools that we are aware
of~\cite{huggingface_pickle,huggingface_protectai,fickling} rely on \texttt{pickletools}
to parse pickle opcodes, which makes discrepancies between \texttt{pickletools}
and the deserialization modules high severity security issues.
We manually investigated all discrepancies that occurred as a result of \texttt{pickletools} raising an error during analysis, but where at least one other module did not.
These discrepancies can be used by attackers to bypass scanners with malicious
pickle objects that deserialize correctly.
Any discrepancy that met this search criteria was considered critical and further
validated by bug bounty feedback during our disclosure process.

\subsubsection{Results}
We identified four discrepancies that produce errors when parsed by \texttt{pickletools}
but not by one or both of the deserializer modules. Details about the discrepancies (\#1, \#2, \#3, and \#5) can be found in~\cref{tab:discrepancy_summary_long}.

\subsubsection{Bug Bounty Case Study and Validation}
To demonstrate the security impact, consider Discrepancy \#1, which is a
mismatch in how numeric arguments are decoded and can lead to pickle scanner
bypasses.
In the \texttt{pickle} and \texttt{\_pickle} deserializer modules, \texttt{INT}
and \texttt{LONG} arguments are treated as base 0 values (and so can be provided
as base 10 or base 16 values).
However, \texttt{pickletools} parses all opcode arguments as base 10 values,
and raises an exception when, \eg a base 16 value like 0x1337 is encountered.
We are able to use this discrepancy to construct a malicious pickle object that,
when scanned by the \texttt{picklescan} tool, does not raise any malicious alerts
in the scan summary (see \cref{fig:poc} in \cref{SEC:APPENDIX}).
However, when the pickle object is loaded, it executes shell commands by
invoking \texttt{posix.system}, a function that is disallowed by pickle
scanners.

We disclosed this discrepancy, along with the other security-relevant
discrepancies described above, and the proof-of-concept exploit to the
\href{https://huntr.com}{huntr.com} bug bounty platform as part of their
open-source bug bounty program in bypassing model scanning tools.
The discrepancies were acknowledged, and we were rewarded with a \$750 bounty
for reporting the vulnerabilities.

\summarybox{\textbf{RQ2 Summary:} Of the \totalErrorsFuzzer discrepancies found, four were security relevant and allowed an attacker to bypass scanners. \tool is able to identify these impactful discrepancies, and they were validated by a bug bounty award.}

\subsection{RQ3: Search Efficiency}\label{SEC:EVALUATION_SEARCH_EFFICIENCY}

\tool is designed to identify discrepancies more efficiently than existing
approaches.
Today, there are no comparable tools for finding errors in pickle
implementations;
  \tool is the first automated tool for detecting implementation
  errors in pickle deserializers and parsers.
Therefore, we cannot compare \tool to an existing baseline tool.
Instead, we compare its ability to identify bugs to a baseline of manual
security analysis.

\subsubsection{Methods}
To create a baseline of manual analysis, we had one researcher with experience
auditing Python source code applications spend approximately 60 person-hours
conducting manual analysis of the three test pickle implementations.
The researcher had no prior knowledge of the \tool-discovered results.
We recorded the discrepancies that the researcher discovered at the end of the
60 hour period.

We compare the discrepancies discovered manually by the researcher with the
discrepancies discovered automatically by \tool, and the time that it took
\tool to discover them.

\subsubsection{Results}

Over the 60 hours of manual analysis, the security researcher discovered
\totalErrorsManual total discrepancies, which were roughly evenly distributed
in time (\ie one was discovered approximately every four hours).
\totalErrorsOverlap of the discrepancies were identical to ones discovered by
\tool during its two fuzzing campaigns.
The other \totalErrorsManualOnly discrepancies were not found by \tool within
the allotted time.
Additionally, there were \totalErrorsFuzzerOnly discrepancies found by \tool
that were missed during manual analysis.
A description of each discrepancy and its associated CPython GitHub issue number is found in Table~\ref{tab:discrepancy_summary_long}.
Of the four discrepancies that \tool \textit{failed} to find, two discrepancies (\#7, \#9) are the result of mishandling large integer arguments to opcodes, and \tool failed to randomly generate these large values.
The remaining two (\#15, \#16) are related to frame handling, and are reached when the relationship between a specific opcode argument and the overall size of the pickle program satisfy a precise condition, which \tool has approximately a $\frac{1}{2^{56}}$ chance of generating randomly.

Despite discovering the same \textit{number} of discrepancies,
\tool discovered all discrepancies in the first 10 minutes.
This is promising for scaling \tool to new, untested pickle implementations as
well for quickly auditing future versions, but it may also indicate that \tool is finding ``shallow'' bugs and then failing to make progress.
We discuss the potential for future work to improve the fuzzing approach in \cref{SEC:DISCUSSION_ALTERNATE}.

\summarybox{\textbf{RQ3 Summary:} In 60 hours of manual analysis, a security auditor discovered \totalErrorsManual discrepancies, the same number of discrepancies that \tool discovered in 10 minutes. \tool is efficient, but failed to identify four of the discrepancies discovered by manual analysis.}

\section{Discussion}\label{SEC:DISCUSSION}

\subsection{Generalizability}\label{SEC:DISCUSSION_GENERALIZABILITY}

\tool presently targets parsing and deserialization discrepancies in just the
three \textit{de facto} CPython pickle modules, but its design allows additional
implementations to be integrated with minimal engineering effort.
To incorporate another pickle deserializer or disassembler, the implementation
must be able to expose the internal storage state in a form suitable for string
comparison and indicate whether an error was raised during the disassembly or
execution of the pickle program, as described in~\cref{SEC:OVERVIEW_DESIGN}.
This enables \tool to evaluate new pickle implementations at essentially no
additional cost, while operating at a speed far exceeding manual analysis.

In practice, we found that the core CPython pickle modules are the most widely
used implementations, with few actively maintained alternatives.
For example, Jython---a Python implementation that runs on the Java
platform---includes a pickle deserialization module~\cite{jython_docs}, but only
supports up to Python 2.7~\cite{jython}, which reached end-of-life in January
2020~\cite{python_eol}.
Other known implementations are either outdated or see little real-world
use~\cite{ironpython,pypy}.
If, in the future, a new implementation gains popularity and requires fast
testing, \tool will be able to compare its executions as long as its internal
storage and error status is accessible.

\subsection{Alternative Fuzzing Strategies}\label{SEC:DISCUSSION_ALTERNATE}

\tool currently employs a grammar-based generational strategy that produces
syntactically valid pickle programs derived from the behavior of the reference
implementations.
Opcodes and their arguments are selected randomly and exercised without runtime
feedback, resulting in broad but unguided exploration.
This black-box approach is effective for quickly surfacing inconsistencies
across implementations, but it does not prioritize inputs that expand code
coverage or stress rarely used argument combinations.
A feedback-driven or coverage-guided fuzzer could bias generation toward
previously unexplored execution paths, potentially increasing the density of
meaningful test cases and reducing time spent on redundant states.

Mutation-based fuzzing offers a complementary avenue. By mutating inputs that
already trigger new coverage or behavioral discrepancies, the fuzzer could
iteratively refine interesting samples and probe deeper into edge cases.
Unlike the current grammar-constrained generator, a mutation strategy need not
enforce strict syntactic validity, which may expose parser differentials or
error-handling inconsistencies when implementations process malformed pickle
streams.

Adopting a hybrid design---grammar-based seeding to ensure semantic
reachability, followed by coverage-guided mutation---may provide the advantages
of both approaches.
The existing method is well suited for establishing a baseline, but more
adaptive techniques are likely to discover deeper or more complex divergences by
concentrating effort on inputs that noticeably influence execution.
Our evaluation (\S\ref{SEC:EVALUATION}) shows that \tool is able to identify
security-relevant discrepancies with its current fuzzing approach, and future
work can explore alternate fuzzing techniques for improved pickle parser
coverage.

\subsection{Responsible Disclosure and Open Source Engagement}\label{SEC:DISCUSSION_RESPONSIBLE_DISCLOSURE}

All discrepancies discovered during this research were disclosed to the Python
Software Foundation via issues and pull requests filed in the CPython
GitHub repository.
Each report included detailed descriptions of root causes and example payloads
demonstrating the inconsistencies across implementations.

Since Python pickle has an implementation-defined specification and limited
documentation regarding individual opcode behavior, it is often unclear which
implementation should be modified to resolve a discrepancy.
In proposed fixes, care was taken to align with any existing documentation or
relevant Python Enhancement Proposals (PEPs).
In cases where no authoritative reference existed, the findings helped initiate
discussion among Python developers to determine the most appropriate resolution.

While some of the discrepancies could be fixed with simple, low-cost changes, others were more complicated or led to additional checks that the maintainers did not feel were worth implementing.
Additionally, some issues have not yet been addressed due to the large number of issues and pull requests the CPython repository receives.

All discrepancies with potential security impact were reported through the huntr.com bug bounty program under its Model File Vulnerability category. The submission was accepted and awarded \$750, though the report visibility remains private at the time of publication.

See Table~\ref{tab:discrepancy_summary_long} for links to each GitHub issue and
current resolution status as of time of publication.

\section{Conclusion}

In this work, we show that differential testing is a successful approach for identifying impactful security discrepancies in pickle modules.
We studied the specification-less pickle implementations to create a grammar, and then used that grammar to implement \tool, the first automated testing tool for pickle implementation correctness.
\tool demonstrates the viability of differential testing for preventing security-critical issues in the pickle modules, avoiding potential future pickle scanner bypasses.
We evaluated \tool on three commonly used pickle implementations:
\texttt{pickle}, \texttt{\_pickle}, and \texttt{pickletools}.
\tool discovered and disclosed \totalErrorsFuzzer discrepancies, four of which were security related.
The findings reveal that inconsistent behavior often exists for the same
protocol across different implementations and codebases and demonstrates the
security impact discrepancies can have in the real world. 
We show that
differential fuzzing contributes to aligning these implementations by
automatically discovering these discrepancies with minimal manual effort.
We hope that \tool continues to provide use to the community by
aiding in differential testing, security auditing, and future pickle-related
development.

\section*{Availability}

\tool's source code is publicly available at GitHub:
\url{https://github.com/Legoclones/PickleFuzzer}.

\section*{acknowledgements}\label{SEC:ACKNOWLEDGEMENTS}

We thank the anonymous reviewers for their constructive feedback on our work.
We also thank Ben Kallus, for inspiring discussion about differential
testing, and James C. Davis, Berk \c{C}akar, Neophytos Christou, Jeffrey Goeders, and Albert Tay for feedback on
drafts of this work.

\bibliographystyle{IEEEtran}
\bibliography{main}

\begin{thebibliography}{10}
\providecommand{\url}[1]{#1}
\csname url@samestyle\endcsname
\providecommand{\newblock}{\relax}
\providecommand{\bibinfo}[2]{#2}
\providecommand{\BIBentrySTDinterwordspacing}{\spaceskip=0pt\relax}
\providecommand{\BIBentryALTinterwordstretchfactor}{4}
\providecommand{\BIBentryALTinterwordspacing}{\spaceskip=\fontdimen2\font plus
\BIBentryALTinterwordstretchfactor\fontdimen3\font minus
  \fontdimen4\font\relax}
\providecommand{\BIBforeignlanguage}[2]{{%
\expandafter\ifx\csname l@#1\endcsname\relax
\typeout{** WARNING: IEEEtran.bst: No hyphenation pattern has been}%
\typeout{** loaded for the language `#1'. Using the pattern for}%
\typeout{** the default language instead.}%
\else
\language=\csname l@#1\endcsname
\fi
#2}}
\providecommand{\BIBdecl}{\relax}
\BIBdecl

\bibitem{slaviero_sensepost_2010}
\BIBentryALTinterwordspacing
M.~Slaviero, ``{SensePost} {\textbar} {Playing} with python pickle \#1,'' Oct.
  2010,
  \url{https://sensepost.com/blog/2010/playing-with-python-pickle-\%231/}.
  [Online]. Available:
  \url{https://sensepost.com/blog/2010/playing-with-python-pickle-\%231/}
\BIBentrySTDinterwordspacing

\bibitem{slaviero_shellcoding_nodate}
\BIBentryALTinterwordspacing
------, ``\BIBforeignlanguage{en}{Shellcoding in {Python}'s serialisation
  format},'' \emph{\BIBforeignlanguage{en}{Black Hat}}, 2011. [Online].
  Available:
  \url{https://media.blackhat.com/bh-us-11/Slaviero/BH_US_11_Slaviero_Sour_Pickles_WP.pdf}
\BIBentrySTDinterwordspacing

\bibitem{pickle-fuzz}
\BIBentryALTinterwordspacing
A.~Willmer, ``moreati/pickle-fuzz,''
  \url{https://github.com/moreati/pickle-fuzz}. [Online]. Available:
  \url{https://github.com/moreati/pickle-fuzz}
\BIBentrySTDinterwordspacing

\bibitem{pep307}
\BIBentryALTinterwordspacing
G.~van Rossum, ``\BIBforeignlanguage{en}{{PEP} 307 – {Extensions} to the
  pickle protocol {\textbar} peps.python.org},'' Feb. 2003,
  \url{https://peps.python.org/pep-0307/}. [Online]. Available:
  \url{https://peps.python.org/pep-0307/}
\BIBentrySTDinterwordspacing

\bibitem{docs_django}
\BIBentryALTinterwordspacing
``\BIBforeignlanguage{en}{Django's cache framework {\textbar} {Django}
  documentation},'' \url{https://docs.djangoproject.com/en/5.1/topics/cache/}.
  [Online]. Available:
  \url{https://docs.djangoproject.com/en/5.1/topics/cache/}
\BIBentrySTDinterwordspacing

\bibitem{sultanik_dill_2021}
\BIBentryALTinterwordspacing
E.~Sultanik, ``\BIBforeignlanguage{en}{Never a dill moment: {Exploiting}
  machine learning pickle files},'' Mar. 2021,
  \url{https://blog.trailofbits.com/2021/03/15/never-a-dill-moment-exploiting-machine-learning-pickle-files/}.
  [Online]. Available:
  \url{https://blog.trailofbits.com/2021/03/15/never-a-dill-moment-exploiting-machine-learning-pickle-files/}
\BIBentrySTDinterwordspacing

\bibitem{kellas_pickleball}
\BIBentryALTinterwordspacing
A.~D. Kellas, N.~Christou, W.~Jiang, P.~Li, L.~Simon, Y.~David, V.~P. Kemerlis,
  J.~C. Davis, and J.~Yang, ``Pickleball: Secure deserialization of
  pickle-based machine learning models,'' in \emph{Proceedings of the 2025 ACM
  SIGSAC Conference on Computer and Communications Security}, ser. CCS
  '25.\hskip 1em plus 0.5em minus 0.4em\relax New York, NY, USA: Association
  for Computing Machinery, 2025, p. 3341–3355. [Online]. Available:
  \url{https://doi.org/10.1145/3719027.3765037}
\BIBentrySTDinterwordspacing

\bibitem{splunk}
\BIBentryALTinterwordspacing
``\BIBforeignlanguage{en}{Paws in the {Pickle} {Jar}: {Risk} \& {Vulnerability}
  in the {Model}-sharing {Ecosystem}},''
  \url{https://www.splunk.com/en_us/blog/security/paws-in-the-pickle-jar-risk-vulnerability-in-the-model-sharing-ecosystem.html}.
  [Online]. Available:
  \url{https://www.splunk.com/en_us/blog/security/paws-in-the-pickle-jar-risk-vulnerability-in-the-model-sharing-ecosystem.html}
\BIBentrySTDinterwordspacing

\bibitem{modelscan}
\BIBentryALTinterwordspacing
``protectai/modelscan,'' Apr. 2025,
  \url{https://github.com/protectai/modelscan}. [Online]. Available:
  \url{https://github.com/protectai/modelscan}
\BIBentrySTDinterwordspacing

\bibitem{picklescan}
\BIBentryALTinterwordspacing
M.~Maitre, ``mmaitre314/picklescan,'' Apr. 2025,
  \url{https://github.com/mmaitre314/picklescan}. [Online]. Available:
  \url{https://github.com/mmaitre314/picklescan}
\BIBentrySTDinterwordspacing

\bibitem{fickling}
\BIBentryALTinterwordspacing
``trailofbits/fickling,'' Feb. 2026,
  \url{https://github.com/trailofbits/fickling}. [Online]. Available:
  \url{https://github.com/trailofbits/fickling}
\BIBentrySTDinterwordspacing

\bibitem{huggingface_protectai}
\BIBentryALTinterwordspacing
``Hugging face teams up with protect ai: Enhancing model security for the ml
  community,'' \url{https://huggingface.co/blog/protectai}. [Online].
  Available: \url{https://huggingface.co/blog/protectai}
\BIBentrySTDinterwordspacing

\bibitem{huggingface_protectai_docs}
\BIBentryALTinterwordspacing
``Third-party scanner: Protect ai,''
  \url{https://huggingface.co/blog/protectai}. [Online]. Available:
  \url{https://huggingface.co/blog/protectai}
\BIBentrySTDinterwordspacing

\bibitem{huggingface_pickle}
\BIBentryALTinterwordspacing
``Pickle {Scanning},''
  \url{https://huggingface.co/docs/hub/en/security-pickle}. [Online].
  Available: \url{https://huggingface.co/docs/hub/en/security-pickle}
\BIBentrySTDinterwordspacing

\bibitem{docs_python_pickle}
\BIBentryALTinterwordspacing
``\BIBforeignlanguage{en}{pickle — {Python} object serialization},''
  \url{https://docs.python.org/3/library/pickle.html}. [Online]. Available:
  \url{https://docs.python.org/3/library/pickle.html}
\BIBentrySTDinterwordspacing

\bibitem{reversinglabs:2025:brokenpicklebypass}
\BIBentryALTinterwordspacing
K.~Zanki, ``Malicious ml models discovered on hugging face platform,'' Feb
  2025. [Online]. Available:
  \url{https://www.reversinglabs.com/blog/rl-identifies-malware-ml-model-hosted-on-hugging-face}
\BIBentrySTDinterwordspacing

\bibitem{liu:2025:hideandseek}
\BIBentryALTinterwordspacing
T.~Liu, G.~Meng, P.~Zhou, Z.~Deng, S.~Yao, and K.~Chen, ``{The Art of Hide and
  Seek: Making Pickle-Based Model Supply Chain Poisoning Stealthy Again},''
  2025. [Online]. Available: \url{https://arxiv.org/abs/2508.19774}
\BIBentrySTDinterwordspacing

\bibitem{huggingface}
\BIBentryALTinterwordspacing
``Hugging {Face} – {The} {AI} community building the future.'' May 2025,
  \url{https://huggingface.co/}. [Online]. Available:
  \url{https://huggingface.co/}
\BIBentrySTDinterwordspacing

\bibitem{python_commit_16c8}
\BIBentryALTinterwordspacing
``cpython/{Lib}/pickle.py at 16c8eccfcf85811d1d9368aacb94b47ae8195719 ·
  python/cpython,''
  \url{https://github.com/python/cpython/blob/16c8eccfcf85811d1d9368aacb94b47ae8195719/Lib/pickle.py#L658}.
  [Online]. Available:
  \url{https://github.com/python/cpython/blob/16c8eccfcf85811d1d9368aacb94b47ae8195719/Lib/pickle.py#L658}
\BIBentrySTDinterwordspacing

\bibitem{huang_pain_2022}
\BIBentryALTinterwordspacing
N.-J. Huang, C.-J. Huang, and S.-K. Huang, ``Pain {Pickle}: {Bypassing}
  {Python} {Restricted} {Unpickler} for {Automatic} {Exploit} {Generation},''
  in \emph{2022 {IEEE} 22nd {International} {Conference} on {Software}
  {Quality}, {Reliability} and {Security} ({QRS})}, Dec. 2022, pp. 1079--1090,
  iSSN: 2693-9177. [Online]. Available:
  \url{https://ieeexplore.ieee.org/abstract/document/10062403}
\BIBentrySTDinterwordspacing

\bibitem{weights_only_unpickler}
PyTorch, ``Weights-only unpickler,''
  \url{https://github.com/pytorch/pytorch/blob/main/torch/_weights_only_unpickler.py},
  2024.

\bibitem{python_github_picklepy}
\BIBentryALTinterwordspacing
``\BIBforeignlanguage{en}{cpython/{Lib}/pickle.py at main · python/cpython},''
  \url{https://github.com/python/cpython/blob/main/Lib/pickle.py}. [Online].
  Available: \url{https://github.com/python/cpython/blob/main/Lib/pickle.py}
\BIBentrySTDinterwordspacing

\bibitem{python_github_picklec}
\BIBentryALTinterwordspacing
``\BIBforeignlanguage{en}{cpython/{Modules}/\_pickle.c at main ·
  python/cpython},''
  \url{https://github.com/python/cpython/blob/main/Modules/_pickle.c}.
  [Online]. Available:
  \url{https://github.com/python/cpython/blob/main/Modules/_pickle.c}
\BIBentrySTDinterwordspacing

\bibitem{python_github_pickletools}
\BIBentryALTinterwordspacing
``\BIBforeignlanguage{en}{cpython/{Lib}/pickletools.py at main ·
  python/cpython},''
  \url{https://github.com/python/cpython/blob/main/Lib/pickletools.py}.
  [Online]. Available:
  \url{https://github.com/python/cpython/blob/main/Lib/pickletools.py}
\BIBentrySTDinterwordspacing

\bibitem{McKeeman1998DifferentialTF}
\BIBentryALTinterwordspacing
W.~M. McKeeman, ``Differential testing for software,'' \emph{Digit. Tech. J.},
  vol.~10, pp. 100--107, 1998. [Online]. Available:
  \url{https://api.semanticscholar.org/CorpusID:14018070}
\BIBentrySTDinterwordspacing

\bibitem{asfuzzer}
\BIBentryALTinterwordspacing
H.~Kim, S.~Kim, J.~Lee, and S.~K. Cha, ``Asfuzzer: Differential testing of
  assemblers with error-driven grammar inference,'' in \emph{Proceedings of the
  33rd ACM SIGSOFT International Symposium on Software Testing and Analysis},
  ser. ISSTA 2024.\hskip 1em plus 0.5em minus 0.4em\relax New York, NY, USA:
  Association for Computing Machinery, 2024, p. 1099–1111. [Online].
  Available: \url{https://doi.org/10.1145/3650212.3680345}
\BIBentrySTDinterwordspacing

\bibitem{kallus_dippygram}
\BIBentryALTinterwordspacing
B.~Kallus and S.~W. Smith, ``dippy\_gram: Grammar-aware, coverage-guided
  differential fuzzing of url parsers,'' 2023. [Online]. Available:
  \url{https://langsec.org/spw23/papers/Kallus_LangSec23.pdf}
\BIBentrySTDinterwordspacing

\bibitem{andarzian:2025:mimedifferentialfuzzing}
\BIBentryALTinterwordspacing
S.~B. Andarzian, M.~Meyers, and E.~Poll, ``{ Email Smuggling with Differential
  Fuzzing of MIME Parsers },'' in \emph{2025 IEEE Security and Privacy
  Workshops (SPW)}.\hskip 1em plus 0.5em minus 0.4em\relax Los Alamitos, CA,
  USA: IEEE Computer Society, May 2025, pp. 26--37. [Online]. Available:
  \url{https://doi.ieeecomputersociety.org/10.1109/SPW67851.2025.00007}
\BIBentrySTDinterwordspacing

\bibitem{equivocal_urls}
\BIBentryALTinterwordspacing
J.~Reynolds, A.~Bates, and M.~Bailey, ``{Equivocal URLs: Understanding the
  Fragmented Space of URL Parser Implementations},'' in \emph{Computer Security
  – ESORICS 2022: 27th European Symposium on Research in Computer Security,
  Copenhagen, Denmark, September 26–30, 2022, Proceedings, Part III}.\hskip
  1em plus 0.5em minus 0.4em\relax Berlin, Heidelberg: Springer-Verlag, 2022,
  p. 166–185. [Online]. Available:
  \url{https://doi.org/10.1007/978-3-031-17143-7_9}
\BIBentrySTDinterwordspacing

\bibitem{pickledoc}
\BIBentryALTinterwordspacing
``\BIBforeignlanguage{en}{pickledoc/{Opcodes.md} at main ·
  {Legoclones}/pickledoc},''
  \url{https://github.com/Legoclones/pickledoc/blob/main/Opcodes.md}. [Online].
  Available: \url{https://github.com/Legoclones/pickledoc/blob/main/Opcodes.md}
\BIBentrySTDinterwordspacing

\bibitem{python_issue_cpickle}
\BIBentryALTinterwordspacing
``\BIBforeignlanguage{en}{{cPickle} doesn't raise error, pickle does
  (recursiondepth) · {Issue} \#38614 · python/cpython},''
  \url{https://github.com/python/cpython/issues/38614}. [Online]. Available:
  \url{https://github.com/python/cpython/issues/38614}
\BIBentrySTDinterwordspacing

\bibitem{mallissery_demystify_2023}
\BIBentryALTinterwordspacing
S.~Mallissery and Y.-S. Wu, ``Demystify the fuzzing methods: A comprehensive
  survey,'' \emph{ACM Comput. Surv.}, vol.~56, no.~3, Oct. 2023. [Online].
  Available: \url{https://doi.org/10.1145/3623375}
\BIBentrySTDinterwordspacing

\bibitem{salls_token-level_2021}
\BIBentryALTinterwordspacing
C.~Salls, C.~Jindal, J.~Corina, C.~Kruegel, and G.~Vigna, ``{Token-Level}
  fuzzing,'' in \emph{30th USENIX Security Symposium (USENIX Security
  21)}.\hskip 1em plus 0.5em minus 0.4em\relax USENIX Association, Aug. 2021,
  pp. 2795--2809. [Online]. Available:
  \url{https://www.usenix.org/conference/usenixsecurity21/presentation/salls}
\BIBentrySTDinterwordspacing

\bibitem{jython_docs}
\BIBentryALTinterwordspacing
``Chapter 5: {Input} and {Output} — {Definitive} {Guide} to {Jython} latest
  documentation,'' \url{https://jython.readthedocs.io/en/latest/InputOutput/}.
  [Online]. Available:
  \url{https://jython.readthedocs.io/en/latest/InputOutput/}
\BIBentrySTDinterwordspacing

\bibitem{jython}
\BIBentryALTinterwordspacing
``\BIBforeignlanguage{en-US}{What is {Jython}},''
  \url{https://www.jython.org/}. [Online]. Available:
  \url{https://www.jython.org/}
\BIBentrySTDinterwordspacing

\bibitem{python_eol}
\BIBentryALTinterwordspacing
``\BIBforeignlanguage{en}{Sunsetting {Python} 2},''
  \url{https://www.python.org/doc/sunset-python-2/}. [Online]. Available:
  \url{https://www.python.org/doc/sunset-python-2/}
\BIBentrySTDinterwordspacing

\bibitem{ironpython}
\BIBentryALTinterwordspacing
``{IronLanguages}/ironpython3,'' Feb. 2026,
  \url{https://github.com/IronLanguages/ironpython3}. [Online]. Available:
  \url{https://github.com/IronLanguages/ironpython3}
\BIBentrySTDinterwordspacing

\bibitem{pypy}
\BIBentryALTinterwordspacing
``pypy/pypy,'' Feb. 2026, \url{https://github.com/pypy/pypy}. [Online].
  Available: \url{https://github.com/pypy/pypy}
\BIBentrySTDinterwordspacing

\end{thebibliography}

\clearpage
\onecolumn
\appendix\label{SEC:APPENDIX}

\section{Proof of Concept Code}
\label{sec:proof_of_concept}

\begin{figure}[h!]
\begin{verbatim}
root@29bc390197e5:~# xxd mal.pkl
00000000: 4930 7831 3333 370a 8c05 706f 7369 788c  I0x1337...posix.
00000010: 0673 7973 7465 6d93 8c06 7768 6f61 6d69  .system...whoami
00000020: 8552 2e                                  .R.
root@29bc390197e5:~# picklescan -p mal.pkl
ERROR: parsing pickle in /root/mal.pkl: invalid literal for int() with base 10: b'0x1337'
----------- SCAN SUMMARY -----------
Scanned files: 0
Infected files: 0
Dangerous globals: 0
root@29bc390197e5:~# python3
Python 3.12.10 (main, May 22 2025, 01:29:12) [GCC 12.2.0] on linux
Type "help", "copyright", "credits" or "license" for more information.
>>> from pickle import _loads # Python
>>> from _pickle import loads # C
>>> mal = open('mal.pkl','rb').read()
>>> _loads(mal)
root
0
>>> loads(mal)
root
0
\end{verbatim}
\caption{\label{fig:poc} Picklescan fails to detect a malicious pickle file
(\texttt{mal.pkl}) even though the file successfully loads and executes a
malicious payload.
This malicious pickle file is manually constructed to leverage Discrepancy \#1
to bypass scanners but invoke a payload consisting of
\texttt{posix.system('whoami')}.
\texttt{xxd} displays the hex contents of this malicious pickle file, showing
the dangerous callables in the payload.
Picklescan scans the file and displays an error, but reports no infected files
or dangerous globals, despite the \texttt{posix.system} callable being
disallowed.
When the pickle object is loaded, using both the C (\texttt{\_pickle}) and
Python (\texttt{pickle}) pickle implementations,
the malicious payload executes.
}
\end{figure}

\end{document}